\documentclass[aps,prl,reprint,amsmath,amssymb,superscriptaddress,showpacs,showkeys]{revtex4-1}

\usepackage{amsmath,amssymb,graphicx}
\usepackage{enumitem}
\usepackage{amsbsy}
\usepackage{latexsym}
\usepackage{color}
\usepackage{graphicx}
\usepackage{psfrag}
\usepackage[normalem]{ulem}
\usepackage{bm}
\usepackage[pdftex,hypertexnames=false]{hyperref}

\newcommand{\be}{\begin{equation}}
\newcommand{\ee}{\end{equation}}
\newcommand{\bea}{\begin{eqnarray}}
\newcommand{\eea}{\end{eqnarray}}

\newcommand{\comment}[1]{}

\begin{document}

\title{Thermodynamic learning}

\author{Federico Corberi}
\email{fcorberi@unisa.it}
\author{Salvatore dello Russo}
\email{sdellorusso@unisa.it}
\author{Giovanni Messuti}
\email{gmessuti@unisa.it}
\author{Silvia Scarpetta}
\email{sscarpetta@unisa.it}
\author{Luca Smaldone}
\email{lsmaldone@unisa.it}
\affiliation{Dipartimento di Fisica ``E.~R. Caianiello'' and INFN, Gruppo Collegato di Salerno,
via Giovanni Paolo II 132, 84084 Fisciano (SA), Italy.}

\begin{abstract}
We discuss the possibility to train a thermodynamic system, whose micro-variables are fully determined through the Hamiltonian by the usual statistical mechanical rules, to perform tasks such as memorization and generalization. Training is achieved by the application of suitable external fields, playing the role of {\it data}. At variance with conventional machine learning, no other logical or algorithmic rules are introduced. The system is amenable, in principle, to exact analytical calculations. We specialize this general approach to a prototypical Ising system with annealed dichotomous couplings and study its learning ability. Results indicate excellent memorization and good generalization capacity already for small systems, and a tendency to improve with system size.    
\end{abstract}

\maketitle

A standard realization of a supervised machine learning (ML) device comprises a series of variables, the neurons, interacting 
via some prescription containing a set of parameters denoted as weights and biases. The particular distribution of the interactions among the neurons defines the structure of the network, typically of a layered type. 
A subset of variables, the input layer, is exposed to labeled data and, due to the interactions among neurons, the feed forward procedure, a result is read on another subset, the output layer. Finally, this result is compared with the label and the level of disagreement, i.e. the value of a cost function, is made as small as possible by tuning the interaction parameters, for instance by means of some gradient descent back propagation technique.  

Statistical mechanics has provided a powerful theoretical framework for understanding supervised learning. Starting from the pioneering analysis of storage capacity and generalization in feed forward neural networks~\cite{Gardner1988, Seung1992, Hertz1991}, statistical-mechanical methods have recently been successfully extended to a broad range of architectures, including modern deep neural networks, providing new theoretical insights into overparameterization, wide flat minima, feature learning, and the geometry of high-dimensional loss landscapes~\cite{10.1093/acprof:oso/9780198570837.001.0001, Zdeborová02092016, Baldassi2020, Baldassi2022, Li2021, Bassetti2025, Pacelli2023, Ricci2025, Lee2025, Aiudi2025, Shan2026}.

Despite these remarkable advances, existing statistical-mechanical approaches are primarily aimed either at analyzing learning algorithms or at designing energy-based architectures whose training still relies on {\it ad-hoc} optimization procedures, such as gradient descent, backpropagation, or contrastive divergence. In all these cases, the microscopic physical dynamics of the system is supplemented by algorithmic rules that are external to the Hamiltonian evolution.
Here we propose a different perspective. Rather than employing statistical mechanics to analyze or optimize learning algorithms, we ask whether supervised learning itself can emerge in a physical system whose degrees of freedom solely obey the thermodynamic principles dictated by statistical mechanics through the Hamiltonian, as found in textbooks for gases and magnets, and static and dynamic properties are determined thereby.
If this was possible, one could imagine to build a real  thermodynamic learning device, without need of computers (we do not deal here with the many and serious practical difficulties this could involve). This will be dubbed {\it thermodynamic learning}.
Our approach opens the way to a different theoretical perspective to understand learning processes entirely based on well-developed techniques of standard statistical mechanics. Exact analytical calculations, in particular, are in principle possible for such a system, as we will discuss further on. 
Despite some similarities with other physically inspired cognitive devices, Boltzmann machines~\cite{Ackley1985BoltzmannMachines, Hinton1986Learning}, to mention a significant example, our approach is 
profoundly different as no tailor-made procedures are adopted in the training 
stage.

\begin{figure}[h]
\vspace{0. cm}
\centering
\rotatebox{0}{\resizebox{.4\textwidth}{!}{\includegraphics{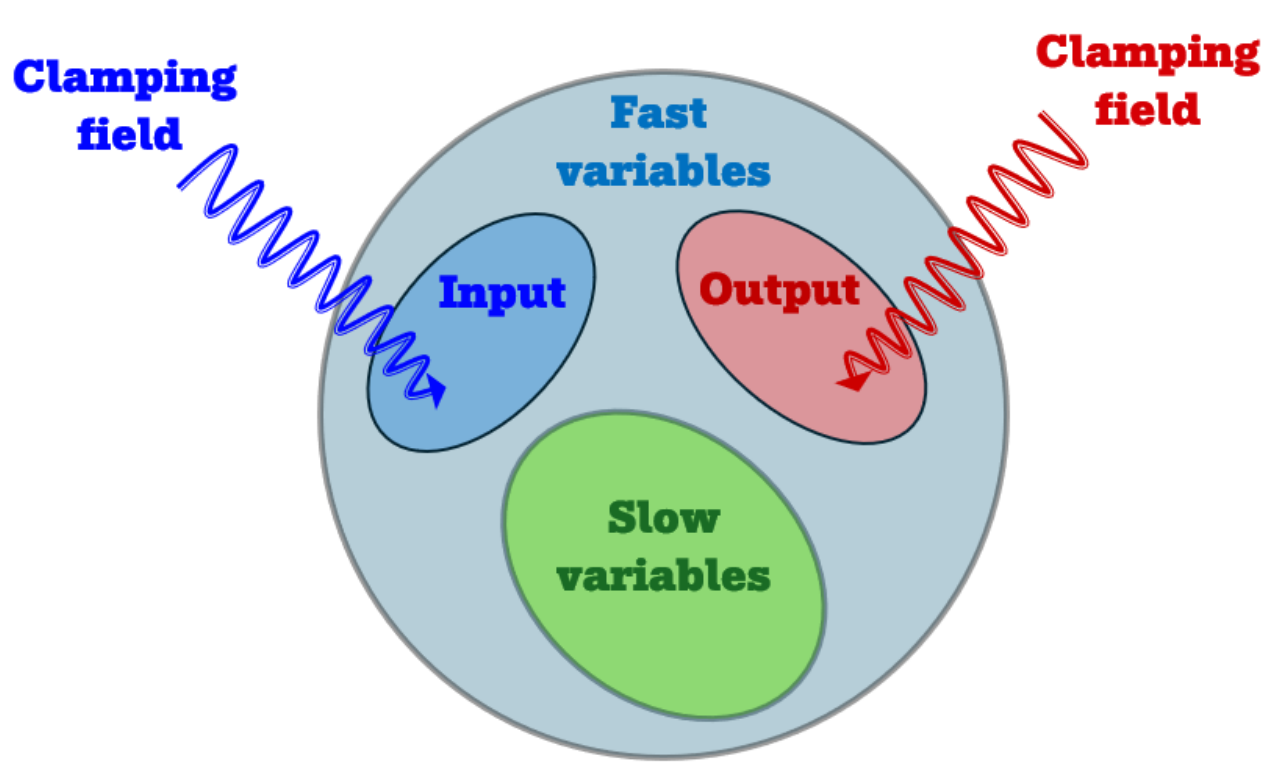}}}
\caption{Schematic drawing of a thermodynamic learning system.}
\label{schema}
\end{figure}

A broad description of the system we have in mind can be given with the help of the schematized illustration  of Fig.~\ref{schema}.
It contains two classes of micro-variables, {\it fast} and {\it slow}, whose typical evolution timescales are widely separated as, for instance, atom vibrations and defects rearrangements in a solid.
Let us denote them as $\{\sigma\}$ and $\{j\}$, respectively. 
Some kind of interaction is present which depends on the $\{\sigma,j\}$ configuration. Fast and slow variables play the role of neurons and weights (and biases) in standard ML, respectively.

Strong external fields, denoted as {\it clamping fields}, interact with two subsets of $\{\sigma\}$, called the input and output sets. They are sufficient to select certain values of the input and output variables,
realizing the {\it clamping}.
Input and output clampings are akin to data and labels, respectively, in usual ML. 

Due to interactions, given a clamping, the unclamped variables will arrange themselves to reach some thermodynamic potential minimum at equilibrium. This is equivalent to the training stage in usual ML.
Then, after quenching the $\{j\}$, namely impeding their further evolution, and removing the output clamping, the output neurons will not hopefully change too much. If this will be the case, we can say that the system has {\it learned} the input field.    

Elaborating further upon this simple idea, we can upgrade the device to learn more than a single input field. This is where evolution timescales come to the forefront. Imagine to change the clampings on an intermediate timescale between the ones of the $\{\sigma\}$ and of the $\{j\}$, so to realize three widely separated timescales. We can foresee that the system will reach a non-equilibrium stationary state where  
the $\{j\}$ will adapt as to match as possible with the clampings, similarly to what was happening before in the equilibrium state in the presence of a single static external field. Having trained the system in this way, we could foresee that, after quenching the $\{j\}$ and removing the output clamping, the output variables will still assume a correct value (i.e. close to the one dictated by the output clamping field in the training stage) for an appreciable subset of the input fields. If this happens, we could again say that the device has {\it learned} the inputs.
In the rest of the article it will be shown, using a specific simple spin model, that thermodynamic learning, as broadly described above, can actually be realized.

\vspace{1pt}

\noindent {\bf A model.}
In order to implement the general idea described above in a particular yet sufficiently general case, we introduce a simple 
statistical mechanics spin model with Boolean variables. Generalizations to other kind of variables are straightforward.

We consider fast variables $\sigma _i=\pm 1$, with $i=1,\dots,N$. Out of these $N$ spins a certain number $N_{in}$, a subset $\{\sigma^{(in)}\}$ denoted {\it input}, is singled out and, similarly, another set $\{\sigma^{(out)}\}$ comprising $N_{out}$ variables, the {\it output}, is selected. The remaining $N_{hid}=N-N_{in}-N_{out}$ spins will be denoted as {\it hidden}.
The slow degrees $j_{ij}=\pm 1$ are also dichotomous, and the interaction between the two kind of variables is ruled by a Hamiltonian term~\cite{N0}
\begin{equation}
{\cal H}_{int}(\{\sigma \},\{j\})=-J\sum _{ij}j_{ij}\sigma_i \sigma_j,
\end{equation} 
where $J$ (and $I$, $L$, below) are interaction strengths.

Bimodal clamping fields $S_i^\alpha=\pm 1$ ($i=1,\dots,N_{in}$), whose number is $n_S$ (i.e. $\alpha=1, \dots, n_S$), compactly denoted as $\{S^\alpha\}$, couple with $\{\sigma^{(in)}\}$ by an Hamiltonian contribution 
\begin{equation}
{\cal H}_{in}(\{\sigma^{(in)}\},\{S^\alpha\})=-I\sum _{i\in in} \sigma_iS_i^\alpha,
\label{Hin}
\end{equation}
where, in the sum, $i$ runs only on the input variables (similarly, on the output variables in Eq.~(\ref{Hout}) below).
$\{S^\alpha\}$ can be thought of as a digital pattern.
To each configuration of the input clampings $\{S^\alpha\}$, output fields $s_i^\alpha=\pm 1$ ($i=1,\dots,N_{out}$) are deterministically associated, which couple to the output neurons through
\begin{equation}
{\cal H}_{out}(\{\sigma^{(out)}\},\{S^\alpha\})=-L\sum _{i\in out} \sigma_is_i^\alpha.
\label{Hout}
\end{equation}
The output fields $\{s^\alpha\}$ play the role of {\it labels} in standard ML and ${\cal H}_{out}$ the one of a loss function.
Notice that ${\cal H}_{out}$ depends, besides on $\{\sigma^{(out)}\}$, only on $\{S^\alpha\}$, because the
$\{s^\alpha\}$ are deterministic functions of the input clampings.
In the limit $I,L\to \infty$, always considered in the following, switching on the clamping fields amounts to impose a {\it boundary} condition on $\{\sigma^{(in)},\sigma^{(out)}\}$. 

Finally, a further contribution to the Hamiltonian is ${\cal H}_S(\{S^\alpha\})$, determining the occurrence of the input fields.
For instance, with reinforced learning, the external resources have their own distribution. In simple cases, in the training stage, a certain set of input data are provided to the system by an operator on equal footing and in this case it is
${\cal H}_S\equiv$ const. $<\infty$ over the training set and ${\cal H}_S=\infty$ otherwise. More in general $p_{S^\alpha}=\frac{e^{-\beta_S{\cal H}_S}}{\sum _\alpha e^{-\beta_S{\cal H}_S}}$ is the {\it a priori} probability of the data.
In conclusion, the full Hamiltonian reads
\begin{equation}
{\cal H}(\{\sigma\},\{j\},\{S^\alpha\})={\cal H}_{int}+{\cal H}_{in}+{\cal H}_{out}+{\cal H}_S.
\label{hamtot}
\end{equation}

Each set of variables $\{j\}$, $\{S^\alpha\}$, $\{\sigma\}$ is attached to thermal baths at inverse temperatures $\beta_j$, $\beta_S$, $\beta_\sigma$, respectively.
The relevant limit is the one $\beta_j,\beta_\sigma \to \infty$, $\beta_S\to 0$, for the following reasons: first, in order to have a good learning $\{j\}$ and $\{\sigma\}$ must relax to the ground state dictated by the clampings, which requires $\beta_j,\beta_\sigma \to \infty$.
Thermal fluctuations are expected to degrade the learning capacity.
Second, for the clampings to be external agents, they must not feel the system variables through the interaction terms (Eqs.~(\ref{Hin},\ref{Hout})), which can only happen for $\beta_S\to 0$. We will compactly indicate this threefold temperature limit as $\bm \beta \to \bm \beta_0$.  

As discussed before, the system is trained by clamping both the input and the output variables and, after that, the latter are set free (setting $L=0$) in the testing stage. 
\vspace{1pt}

\noindent{\bf Analytical approach.}
\noindent{\it Training:}
Borrowing on disordered systems techniques~\cite{mezard1987spin,RWPenney_1993,PhysRevE.62.845,JvanMourik_2001,10.21468/SciPostPhys.10.5.113} a statistical mechanical approach to the problem, whereby analytical calculation can be afforded, is 
arrived at by arguing as follows.
Due to timescales separation, we can imagine the $\{\sigma\}$ have spanned all the possible configurations and reached thermal equilibrium at the actual value of the $\{j,S^\alpha\}$. Hence 
the probability of the $\{\sigma\}$ configurations is expected to be in Boltzmann-Gibbs form $e^{-\beta_\sigma {\cal H}(\{\sigma\},\{j,S^\alpha\})}$.
We can then argue that the quantity $F_{j,S}(\{j,S^\alpha\})$ defined by
\be
Z_{j,S}\equiv e^{-\beta_\sigma F_{j,S}}\equiv \sum _{\{\sigma\}}e^{-\beta_\sigma ({\cal H}_{int}+{\cal H}_{in}+{\cal H}_{out})},
\label{defFjS}
\ee
plays the role of a free energy
(or effective Hamiltonian) 
ruling the behavior of the $\{j,S^\alpha\}$. 
Then the probability of $\{S^\alpha \}$ configurations is proportional to $e^{-\beta _S[F_{j,S}(\{j,S^\alpha\})+{\cal H}_S(\{S^\alpha\})]}$. Repeating the argument, the quantity $F_j(\{j\})$ given by
\be
e^{-\beta_S F_j}\equiv \sum _\alpha e^{-\beta _S (F_{j,S}+{\cal H}_S)}
\label{freeEnj}
\ee
is an effective free energy for the $\{j\}$, meaning that
\begin{equation}
P_j(\{j\})=Z^{-1}e^{-\beta_j F_j},
\label{defpesi}
\end{equation}
where 
\begin{equation}
\begin{split}
	&Z \equiv e^{-\beta_j F}
	= \sum_{\{j\}} e^{-\beta_j F_j} \\
	&= \sum_{\{j\}}
	\left[
	\sum_\alpha e^{-\beta_S {\cal H}_S}
	\left(
	\sum_{\{\sigma\}} e^{-\beta_\sigma
		({\cal H}_{int}+{\cal H}_{in}+{\cal H}_{out})}
	\right)^{n}\,
	\right]^{1/m}
\end{split}
\label{plainZ}
\end{equation}
is a normalization (and we have defined $n=\frac{\beta_S}{\beta_\sigma}$, $m=\frac{\beta_S}{\beta_j}$),
is the probability to observe 
a couplings configuration $\{j\}$ in training (after stationarization has been achieved by all quantities).

Up to this point the parameters $n,m$ are arbitrary. Letting $n=m=1$ one has equilibrium with all the
variables $\{\sigma\}$, $\{S^\alpha\}$ and $\{j\}$ annealed at the same temperature.  
Instead, the limit $\beta_S \to 0$, corresponding to $n,m\to 0$, amounts to a case where the $\{S^\alpha\}$ are seen as {\it quenched}~\cite{RWPenney_1993,PhysRevE.62.845,JvanMourik_2001,10.21468/SciPostPhys.10.5.113}, in the sense of disordered systems, by the $\{\sigma\}$~\cite{N3}. 

It is important to remark that $n,m \to 0$ can be realized in different ways playing with $\beta_\sigma, \beta_S, \beta_j$, and the result one arrives at in Eq.~(\ref{defpesi}) may depend on that. 
As shown in the Supplemental Material (SM)~\cite{SM} (Sec.~S1),  
when $\beta _S\to 0$ is taken after $\beta_j,\beta_\sigma \to \infty$, the $\{j\}$ configurations with an associated finite weight $P_j$ are those (denoted by $\{j_\ll\}$) whereby the lowest possible energy over a SINGLE input field is achieved. 
Instead, when $\beta_S \to 0$ is taken first, $P_j\neq 0$ on another set of $\{j\}$ (denoted by $\{j_<\}$): the ones realizing an overall optimization over ALL the inputs. Physical interpretation of this is rather transparent: in the former case, since $\beta_S \neq 0$ when the other two limits are taken, the inputs can accord to the current {\it belief}, namely the actual state of the network. Then the best optimization over single patterns (chosen according to actual belief) is realized by the system. We will indicate with $\bm \beta \to \bm \beta _0^{post}$ this order of the limits. When the other order is considered, denoted by $\bm \beta \to \bm \beta _0^{prior}$, images cannot harmonize with the system, hence their occurrence is only dictated by the {\it a priori} probability $p_{S^\alpha}$.

\noindent{\it Testing:}
In testing we quench the $\{j\}$ and let $L=0$. Then, we switch on an input field $\{S^\alpha\}$, monitor the output and repeat the procedure for all the $n_S$ inputs.

\noindent -- {\it Memorization:}
The memorization attitude of the device can be monitored by a {\it score}, or {\it fidelity} matrix
\begin{equation}
K_{\alpha \gamma}(\{j\})=\frac{1}{Z_{j,S=\alpha}}\sum _{\{\sigma\}} {\cal K}_{\alpha \gamma}e^{-\beta_\sigma ({{\cal H}_{int}+{\cal H}_{in}})}
\label{scoretest}
\end{equation}
where $Z_{j,S=\alpha}$ is the quantity in Eq.~(\ref{defFjS}) evaluated for pattern $\alpha$, and
\begin{equation}
{\cal K}_{\alpha \gamma}={\cal K}_{\alpha \gamma}(\{\sigma^{(out)}\},\{S^\alpha\},\{S^\gamma\})=\frac{1}{N_{out}}\sum _{i\in out}\sigma_i s_i^\gamma
\label{befcrit}
\end{equation}
is the overlap between the output $\{\sigma ^{(out)}\}$ when the (static) input clamping is $\{S^\alpha\}$ and the label corresponding to the input $\{S^\gamma\}$. We will say that the network is memorizing pattern
$\alpha$ if $K_{\alpha \alpha}> K_{\alpha \gamma}$, $\forall \gamma \neq \alpha$. Notice that
$K_{\alpha \gamma}$ depends on the $\{j\}$ sampled in training with probabilities~(\ref{defpesi}).
A quite complete information on the memorization attitude is given by the couple
$(K_{\alpha \gamma},P_j)$.
An aggregate measure is the {\it average} performance
\begin{equation}
\langle K_{\alpha \gamma}\rangle=\sum _{\{j\}}P_jK_{\alpha \gamma}.
\label{avescoretest}
\end{equation}
Eq.~(\ref{avescoretest}) has the form of a quenched average (over $\{j\}$). However, since the concept of self-averaging is possibly not appropriate here, the meaning of Eq.~(\ref{avescoretest}) is that of an average upon many repetitions of the experiment, stopping the training at (large) random times in order to appropriately sample the $\{j\}$ measure. On the other hand, in practical applications, one could
imagine to use the $\{j\}$ providing the best performance (the ones in Eq.~(S14) of SM~\cite{SM} in the example considered there). 

In the following, when not explicitly stated, we will focus on the limit $\bm \beta \to \bm \beta_0$ where,
by construction, $K_{\alpha \gamma}$ ranges from -1 to 1. Regarding the way this limit is realized,
the learning performance is expected to be 
better along the path $\bm \beta \to \bm \beta_0^{prior}$, where a {\it global}  input optimization is realized, rather than along $\bm \beta \to \bm \beta_0^{post}$. 
Notice that a necessary requirement for $K_{\alpha \alpha}=1, \forall \alpha$, for a certain $\{j\}$ is that, letting $\{\sigma^{(out)}\}=\{s^\alpha\}$ a marginally frustrated configuration 
of the system exists for any input pattern $\alpha$, meaning that each spin has less than half of its couplings unsatisfied. The condition is not sufficient, because other configurations
with $\{\sigma^{(out)}\}\neq \{s^\alpha\}$ can satisfy the requirement with equal, or even smaller $F_{j,S}$.

Clearly, the score depends on the number of $n_S$ of training inputs. Therefore
we will use in the following the symbol $K_{\alpha \gamma}(n_S)$. It is clear that $K_{\alpha \alpha}(1)\to 1$
always for $\bm \beta \to \bm \beta_0$, since at least one pattern can always be
retained, and is expected to decrease for larger $n_S$. 
A related observation is that
$K_{\alpha \gamma}(n_S)$, in general, depends on the actual shape of the particular set of $n_S$ training patterns. With this in mind, in order to have an easier interpretation of the results, 
in the following, we will always consider orthogonal patterns~\cite{N1} (in order to avoid superposition among them) and we will set labels equal to the inputs, $\{s^\alpha\}\equiv \{S^\alpha\}$, $\forall \alpha =1,\dots,n_S$ (implying $N_{in}=N_{out}$).
This amounts to having an internal symmetry among input data, the consequence of which is that $K_{\alpha \alpha}$ is independent of $\alpha$ and $K_{\alpha \gamma}$, with $\gamma \neq \alpha$ is independent of $\alpha, \gamma$~\cite{N5}.   

\noindent -- {\it Generalization:}
Usually one is not only interested in memorization but also in the capability 
to generalize, i.e. the capacity to correctly classify inputs $\widetilde \alpha$ which 
are partly different from the ones, $\alpha $, the network was trained on.
Then we will also consider $K_{\widetilde \alpha \gamma}(n_S,\mu)$, namely the testing fidelity computed with input data $\widetilde \alpha $ obtained from the training ones $\alpha$
by reversing a fraction $\mu$ of the $\{S^\alpha\}$. Notice that using deformed patterns in testing spoils the internal symmetry among inputs, reinstating the full dependence of $K_{\widetilde \alpha \gamma}$
on the indices $\widetilde \alpha, \gamma$.
\vspace{1pt}

\begin{figure}[h]
\vspace{0cm}
\centering
\rotatebox{270}{\resizebox{.3\textwidth}{!}{\includegraphics{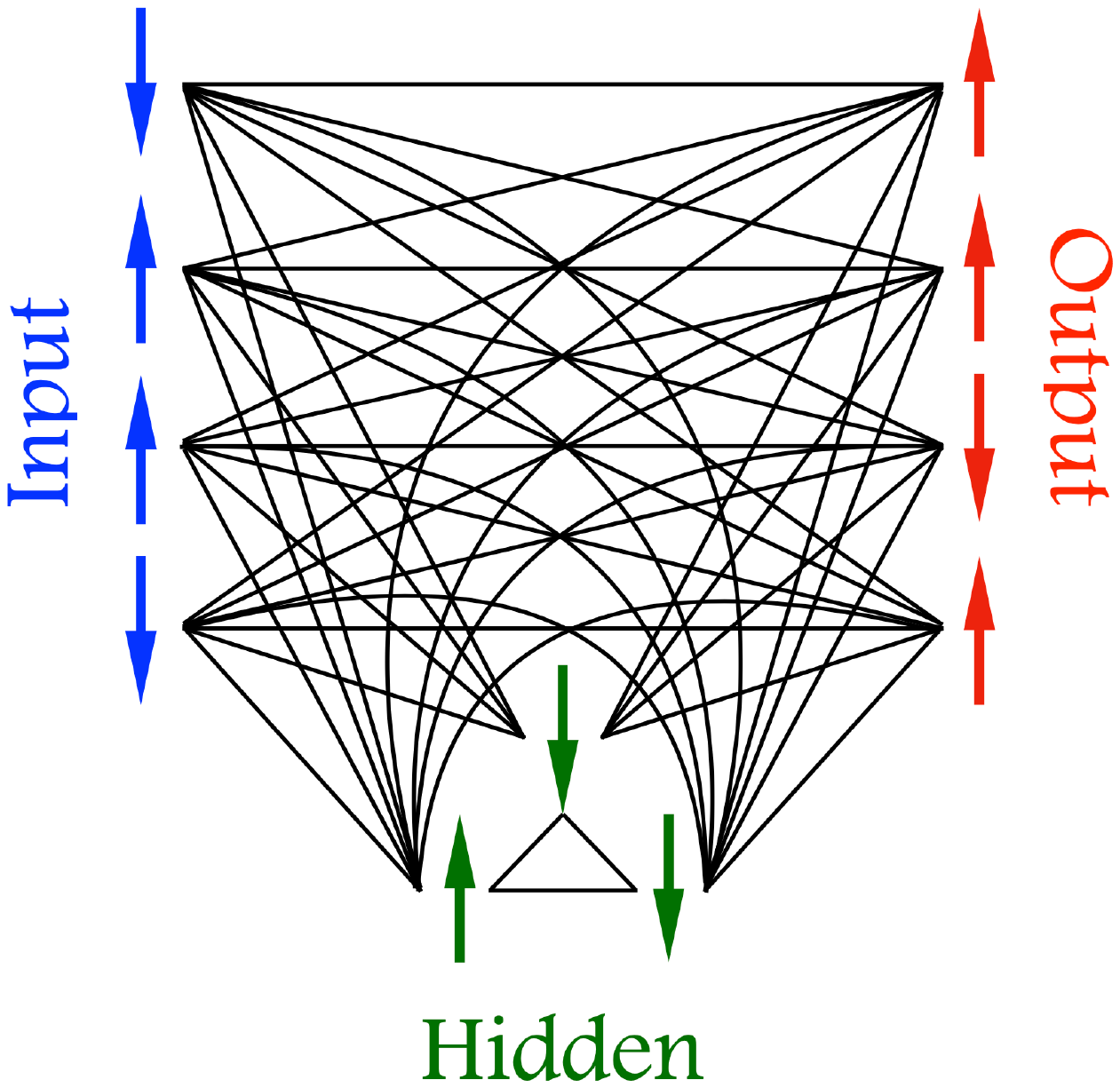}}}
\caption{The all-to-all network considered (for $N_{in}=4, N_{hid}=3, N_{out}=4$).}
\label{fig_net}
\end{figure}

\noindent{\bf A case study.}
In the following we study the learning capacity of an all-to-all network with $N_{in}=N_{out}$, shown in Fig.~\ref{fig_net} for the case denoted $4-3-4$, namely with 
$N_{in}=4, N_{hid}=3, N_{out}=4$~\cite{N7}. We will always consider the most relevant case $\bm \beta \to \bm \beta_0^{prior}$ with $b=\beta_j/\beta_\sigma=1$. Patterns are taken from the Hadamard matrix  
of order $N_{in}$, and hence $N_{in}$ must be 2 or a multiple of 4 and $n_S\le N_{in}$~\cite{N10}. 

According to Eq.~(S6) of the SM~\cite{SM} the task is to determine the 
set $\{j_<\}$ and the associated degeneracies $d$, which boils down to
finding the absolute minima (with respect to $\{j\}$) of the function
$\sum _\alpha {\cal H}_<$, ${\cal H}_<$ being the minimum value (with respect to $\{\sigma\}$) of the Hamiltonian~(\ref{hamtot}) at fixed $\{S^\alpha,j\}$. This task can be done by exact enumeration for sufficiently small architectures or, for larger systems, by Monte Carlo techniques where $\{j\}$ and $\{\sigma\}$ perform  suited stochastic processes. Let us remark that, despite the cardinality of $\{j_<\}$ can be rather large already for small networks, such configurations are hard-to-find remote spots in an immense $\{j\}$ phase space (as size grows), making their scouting hard, similarly to tracking down ground states in spin-glasses.
This explains why in the following we will show results for networks at most large as $16-1-16$ (meaning, however, $2^{288}$ possible $\{j\}$ configurations).

We refer to Sec.~S2 of the SM~\cite{SM} for a complete study of the small prototypical $2-1-2$ network, 
which exhibits in a nutshell many of the features of larger systems, where physical mechanisms can be more easily understood. 
The results for generic networks are reported below. 

\begin{figure}[h]
	\vspace{0cm}
	\centering
	\rotatebox{0}{\resizebox{.5\textwidth}{!}{\includegraphics{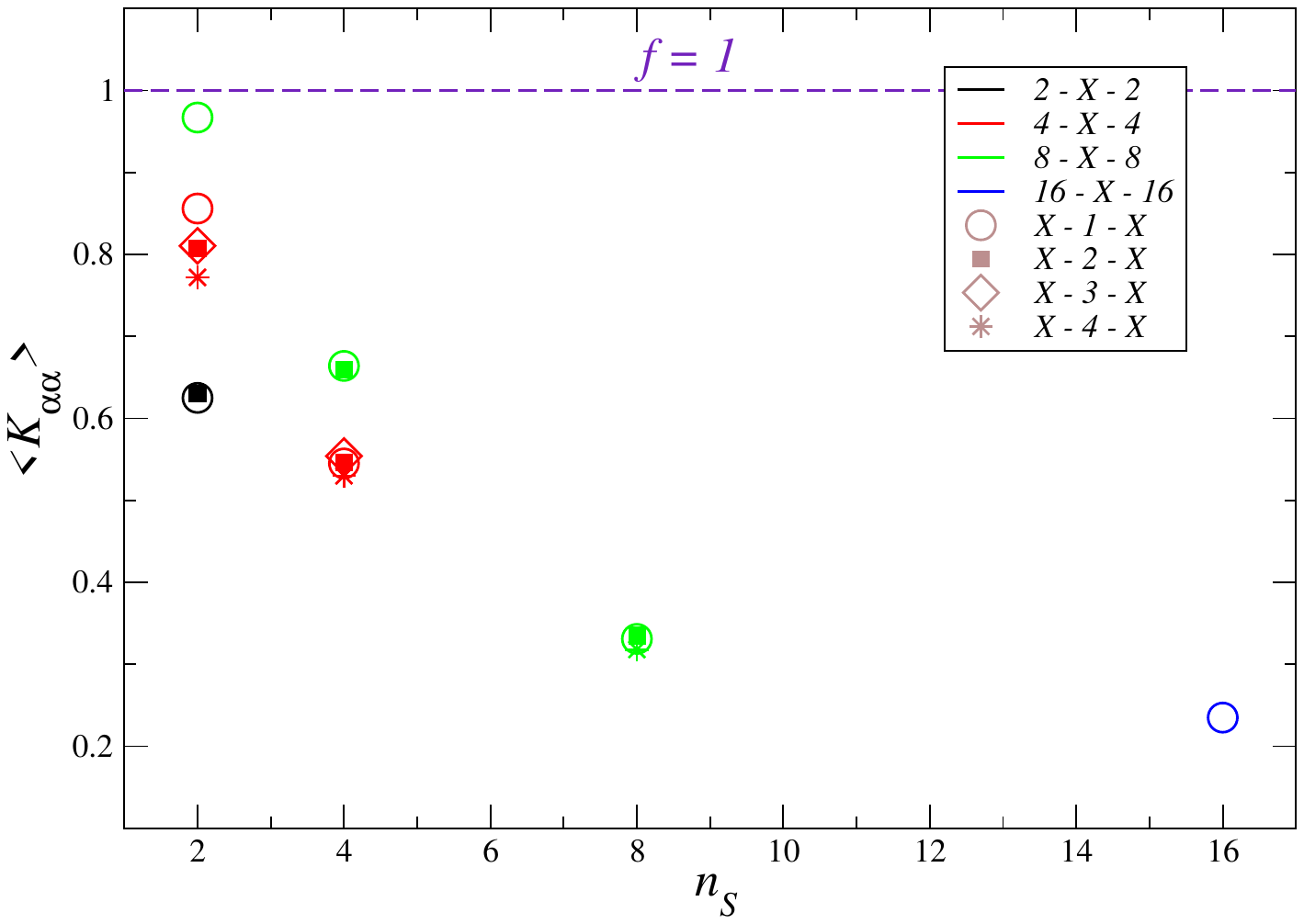}}}
	\rotatebox{0}{\resizebox{.5\textwidth}{!}{\includegraphics{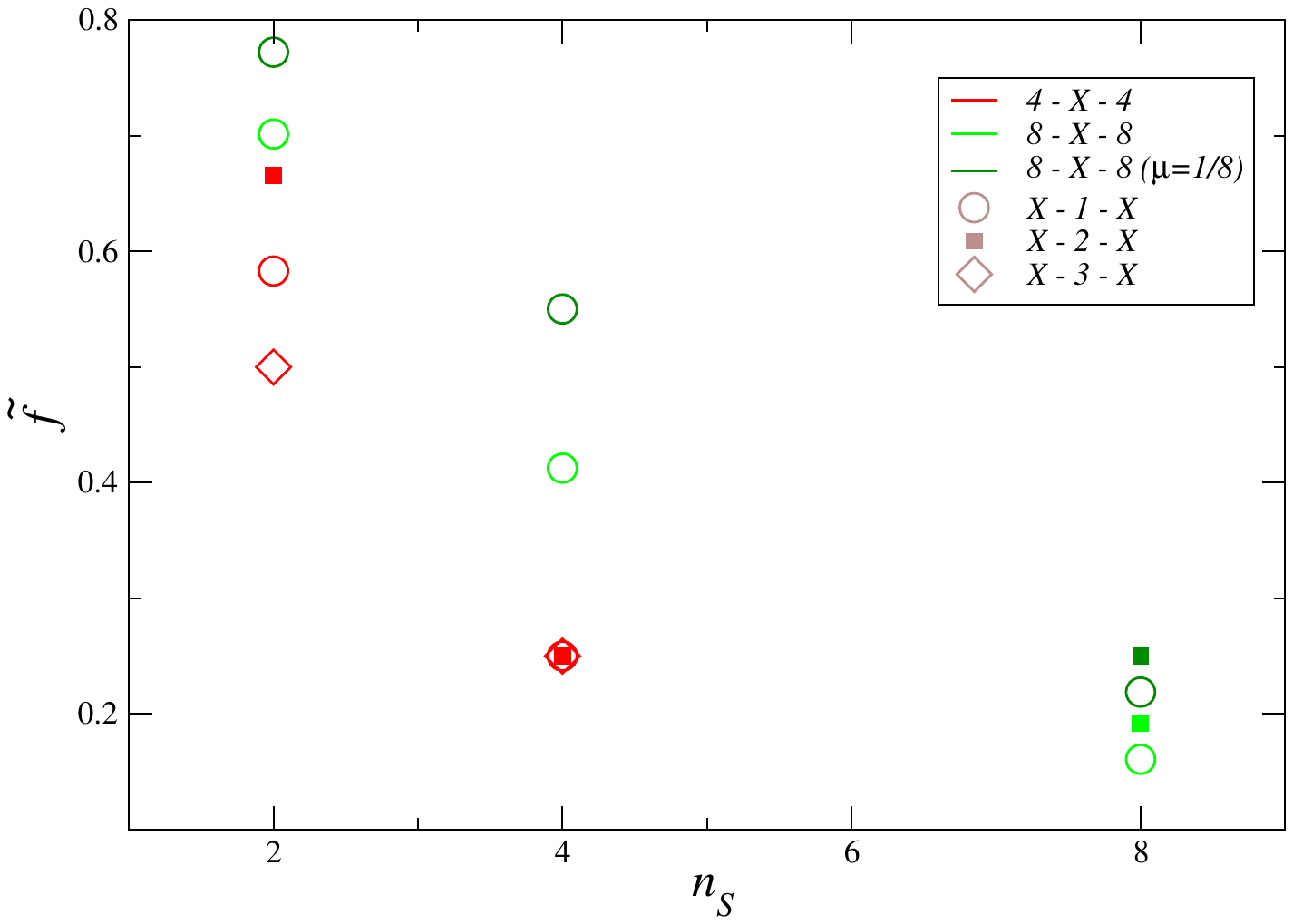}}}
	\caption{Memorization (top) and generalization (bottom) properties of the system. In upper panel $f\equiv 1$ is plotted as a dotted line, and symbols refer to the fidelity values $\langle K_{\alpha \alpha}\rangle $, in lower panel $\widetilde f$ is only plotted. In both panels different colors refer to networks with different $N_{in}$ ($N_{in}=2,4,8,16$) corresponding to, black, red, green, blue, respectively. Given a color, different symbols refer to the value of $N_{hid}$ ($N_{hid}=1,2,3,4$ corresponding to circle, square, diamond, star, respectively). Same colors/symbols are used in the lower part of the figure for $\mu=1/4$. For $\mu =1/8$ and $N_{in}=8$ we use the dark green.}
	\label{fig_mem_gen}
\end{figure}

\noindent{\it Memorization:}
The fidelity of the system is summarized in the upper part of Fig.~\ref{fig_mem_gen} (numerical values are reported in the SM~\cite{SM} (Sec.~S4)). According to the criterion discussed above (below Eq.~(\ref{befcrit})) we can conclude that all the networks considered memorize any possible number of orthogonal patterns~\cite{N1}. Hence, defining $f$ as the fraction of memorized data, we have $f=1$ in any case. This is a first important result of our study. $K_{\alpha \alpha}$, instead, is an indicator of the difficulty, so to say, for the network, to memorize.
Let us stress that $K_{\alpha \alpha}$ can be significantly larger than its average reported in Fig.~\ref{fig_mem_gen} for some of the $\{j_<\}$.
For instance, in the case $4-1-4$, around $50\%$ 
of the $\{j_<\}$ yield $K_{\alpha \alpha}=1$.
At variance with smaller networks, starting from the $4-4-4$ case we have not been able to find all $\{j_<\}$, due to unaffordable long computational time, and the results refer to the limited sample we have detected (see SM~\cite{SM}, Sec. S4). Fig.~\ref{fig_mem_gen} (upper part) suggests some general features: i) For fixed $N_{in}$, $n_S$, changing $N_{hid}$ does not seem to have a major impact for sufficiently large $n_S$ (at least for the restricted range of $N_{hid}$ considered here), as witnessed by the good superposition of different symbols with equal colors for $n_S=4$ and $n_S=8$. The same is true for $n_S=2$ and $N_{in}=2$ (black symbols) whereas, still for $n_S=2$, a certain spread, but without a clear trend, is present only for $N_{in}=4$ (red symbols).
ii) $\langle K_{\alpha \alpha} \rangle$ generally decreases when trying to memorize more patterns, as expected. iii) More interestingly, for given $n_S$, $\langle K_{\alpha \alpha}\rangle$ looks as an  increasing function of $N_{in}$ (see, e.g., the growth of $\langle K_{\alpha \alpha}\rangle$ in going from black, to red and green symbols for $n_S=2$). This suggests that the memorization task is easier in larger systems. 

%

\noindent{\it Generalization:}
In general, for a given input $\alpha $, $K_{\widetilde \alpha, \gamma}$ depends on the choice of $\widetilde \alpha$ (in addition, it also depends on $\alpha$).
We studied the generalization properties by considering, for a certain $\mu\ge 1/4$ (hence $N_{in}\ge 4$) and a given set of $n_S$ patterns $\alpha$, all the possible deformations $\tilde \alpha$~\cite{N6}. In the lower part of Fig.~\ref{fig_mem_gen} the fraction $\widetilde f$ of recognized $\widetilde \alpha$ patterns is displayed (numerical values are reported in the SM~\cite{SM} (Sec.~S4)). Looking at the data for $\mu =1/4$ (i.e. excluding dark-green symbols), this figure shows that properties i), ii) and iii) discussed above regarding memorization seem to remain true also for generalization. In short,
i) dependence on $N_{hid}$ is weak or without a clear trend, ii) generalization gets worse for larger numbers of patterns whereas iii) it increases with $N_{in}$. In addition, one can appreciate that lowering the deformation $\mu$ from $1/4$ to $1/8$ (i.e. going from light to dark-green symbols) increases the performance, as expected.

In this Letter we have introduced and studied a new paradigm of supervised learning, thermodynamic learning. 
In our framework, learning emerges directly from the thermodynamic evolution of the system, fostering  analytical approaches through a statistical-mechanical description.
Being an {\it energy based} approach,
it shares some similarities with the Hopfield model and Boltzmann machines.
The major difference is the training  
procedure which, in the present approach, entirely relies on {\it physical} rules, where separation of timescales is enforced to maximize the learning capacity.
The prototypical spin system considered here exhibits robust memorization and nontrivial generalization, with both showing a tendency to improve with system size. More generally, thermodynamic learning provides a framework for investigating learning as a statistical-mechanical process in its own right and may provide guidelines for realizing standalone, computer-free learning devices. In this respect, our approach complements recent developments in physical learning and physical neural computing, 
which investigate learning directly implemented in physical substrates~\cite{stern2021physnets,stern2023learning,Fischer_2026}. 

\bibliography{biblio_ML}

\end{document}